\documentclass[prl,twocolumn,nofootinbib,floatfix,subeqn]{revtex4}
\usepackage[usenames]{color}
\usepackage{amsmath}
\usepackage{amssymb}
\usepackage{float}
\usepackage{graphicx}
\usepackage{xspace}
\usepackage{hyperref}

\newcommand{\ket}[1]{\ensuremath{\left|#1\right\rangle}}

\newcommand{\alat}{a_{\rm lat}}
\newcommand{\erf}{\mbox{erf}}

\long\def\symbolfootnote[#1]#2{\begingroup%
\def\thefootnote{\fnsymbol{footnote}}\footnotetext[#1]{#2}\endgroup}

\begin{document}

\title{Single-Spin Addressing in an Atomic Mott Insulator}

\author{Christof Weitenberg$^{1}$}
\author{Manuel Endres$^{1}$}
\author{Jacob F. Sherson$^{1\dag}$}
\author{Marc Cheneau$^{1}$}
\author{Peter Schau\ss$^{1}$}
\author{Takeshi Fukuhara$^{1}$}
\author{Immanuel Bloch$^{1,2}$}
\author{Stefan Kuhr$^{1*}$}

\date{9 January 2011}

\affiliation{
   $^1$Max-Planck-Institut f\"ur Quantenoptik, Hans-Kopfermann-Str.~1, 85748 Garching,
   Germany\\
   $^2$Ludwig-Maximilians-Universit\"at, Schellingstr.~4/II, 80799 M\"unchen, Germany
}

\begin{abstract}
Ultracold atoms in optical lattices are a versatile tool to investigate fundamental properties of quantum many body systems. In particular, the high degree of control of experimental parameters has allowed the study of many interesting phenomena such as quantum phase transitions and quantum spin dynamics.
Here we demonstrate how such control can be extended down to the most fundamental level of a single spin at a specific site of an optical lattice.
Using a tightly focussed laser beam together with a microwave field, we were able to flip the spin of individual atoms in a Mott insulator with sub-diffraction-limited resolution, well below the lattice spacing. The Mott insulator provided us with a large two-dimensional array of perfectly arranged atoms, in which we created arbitrary spin patterns by sequentially addressing selected lattice sites after freezing out the atom distribution. We directly monitored the tunnelling quantum dynamics of single atoms in the lattice prepared along a single line and observed that our addressing scheme leaves the atoms in the motional ground state. Our results open the path to a wide range of novel applications from quantum dynamics of spin impurities, entropy transport, implementation of novel cooling schemes, and engineering of quantum many-body phases to quantum information processing.
\end{abstract}

\maketitle

\symbolfootnote[2]{present address: Department of Physics and
Astronomy, University of Aarhus, DK-8000 Aarhus C, Denmark.}
\symbolfootnote[1]{Electronic address: {\bf stefan.kuhr@mpq.mpg.de}}

The ability to observe and control the position of single atoms on a surface of a solid via scanning tunnelling and atomic force microscopy has revolutionised the field of condensed matter physics \cite{Binnig:1987,Giessibl:2003}. In few-atom systems, coherent control of single particles in e.g.~an ion chain has proven crucial for the implementation of high-fidelity quantum gates and the readout of individual qubits in quantum information processing \cite{Blatt:2008}. Bringing such levels of control to the regime of large scale many-body systems has been a longstanding goal in quantum physics. In the context of ultracold atoms in optical lattices, a major challenge has been to combine degenerate atomic samples with single-site addressing resolution and single-atom sensitivity.
This full control is essential for many applications in condensed matter physics, such as the study of spin impurities \cite{Zvonarev:2007} and quantum spin dynamics \cite{Recati:2003,Kleine:2008} within quantum magnetism, entropy transport, the implementation of novel cooling schemes \cite{Weiss:2004,Bernier:2009} or digital quantum simulations based on Rydberg atoms \cite{Weimer:2010}. For scalable quantum information processing, a Mott insulator with unity filling provides a natural quantum register with several hundreds of qubits. In order to exploit the full potential of such a large scale system for quantum computation, coherent manipulation of individual spins is indispensable, both within a circuit-based \cite{Nielsen:2000} or a one-way quantum computer architecture \cite{Raussendorf:2001,Briegel:2009}.

The quest to address atoms on single sites of an optical lattice has
a long history \cite{Dumke:2002,Bergamini:2004,Saffman:2004,Calarco:2004,Weiss:2004,Zhang:2006,Joo:2006,Cho:2007,Gorshkov:2008,Lundblad:2009,Shibata:2009}. In one dimension, single-site addressing was accomplished optically in a long-wavelength lattice \cite{Scheunemann:2000}, in which however tunnelling was completely suppressed, and  using magnetic resonance techniques in a sparsely filled short-wavelength lattice \cite{Schrader:2004,Karski:2010}.
In two dimensions, an electron beam was used to depopulate sites of a Bose-Einstein condensate loaded into an array of potential tubes, each containing up to 80 atoms \cite{Wuertz:2009}. In this case, coherent spin manipulation was not possible and the readout was done by averaging over more than 100 single images.  None of the experiments so far has shown single-atom spin control in strongly correlated systems together  with high fidelity single-atom detection.

Here we report on the achievement of this goal, by demonstrating
single-site-resolved addressing and  control of the spin
states of individual atoms in a Mott insulator in an optical
lattice. The Mott insulator provided us with an almost perfect
initial two-dimensional array of atoms in the same spin state. Apart from few thermal defects, each lattice site contained a single atom in its motional ground
state \cite{Bakr:2010,Sherson:2010}. Using a tightly focused laser
beam, we introduced a controlled differential energy shift between
two atomic spin states at a given lattice site. Microwave radiation
resonant with this shifted transition then allowed us to selectively address
the spin of a single atom \cite{Weiss:2004,Zhang:2006} with high
fidelity. We thereby
obtained sub-diffraction-limited spatial resolution well below the
lattice spacing. By moving the addressing laser beam to different
lattice sites and by inducing spin-flips in the Mott insulator, we
were able to deterministically create arbitrary
two-dimensional spin patterns of individual atoms, thereby realising a
scalable single-atom source \cite{Schlosser:2001,Kuhr:2001,Gruenzweig:2010}.
Furthermore, we investigated how much
our single-spin manipulation affects the motional state of the
atoms by directly monitoring the tunnelling dynamics of single atoms after addressing them. Averaging over several snapshots after
different tunnelling times, we fully reconstructed the
characteristic spatial probability distribution of the single-atom
wave function and its coherent evolution over more than 20 lattice sites.
We were able to discriminate the dynamics of the atoms in the zeroth and in the first band and found that most of the atoms remained in the motional ground state after addressing.



%
\begin{figure}[!t]
    \begin{center}
        \includegraphics[width=\columnwidth]{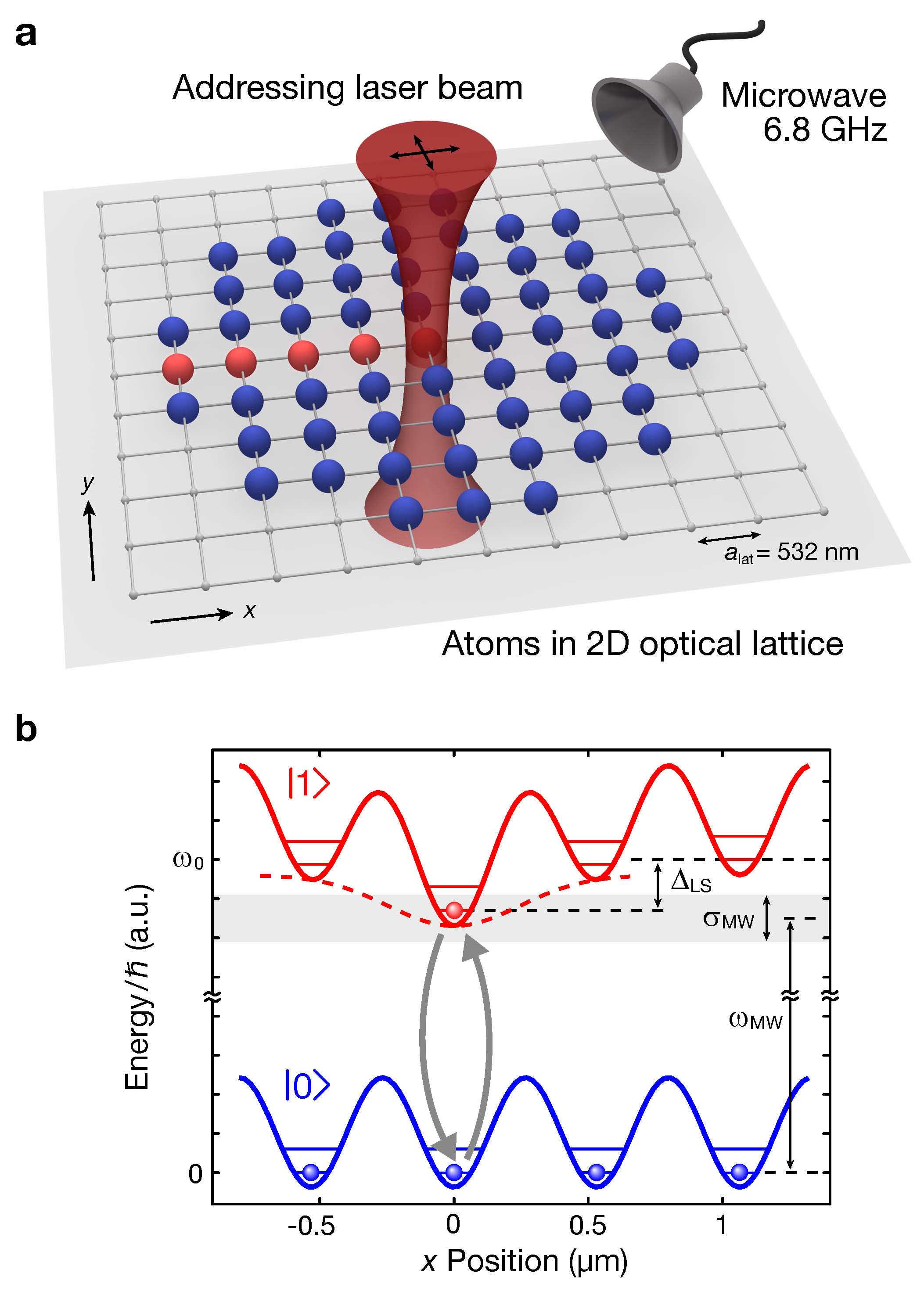}
     \end{center}
     \vspace{-0.5cm}
    \caption{{\bf Addressing scheme.} {\bf a,} Atoms in a Mott insulator
with unity filling arranged on a square lattice with period
$\alat=532$\,nm were addressed using an off-resonant laser beam. The
beam was focussed onto individual lattice sites by a high-aperture
microscope objective (not shown) and could be moved in the $xy$ plane with an accuracy of better than $0.1\,\alat$.  {\bf b,} Energy
diagram of atoms in the lattice for the two hyperfine states
$\ket{0}$ and $\ket{1}$. The $\sigma^-$-polarized addressing beam
locally induces a light shift $\Delta_{\rm LS}$ of state $\ket{1}$,
bringing it into resonance with a microwave field. A Landau-Zener sweep (central frequency $\omega_{\rm MW}$, sweep width $\sigma_{\rm MW}$)
transfers the addressed atoms from $\ket{0}$ to
$\ket{1}$.\label{fig:schematic}}
\end{figure}

\section*{Experimental setup}
In our experiments, we prepared a two-dimensional sample of  ultracold
$^{87}$Rb atoms in an optical lattice, confined in a single antinode
of a vertical standing wave along the $z$ direction. Two pairs of counterpropagating laser
beams (wavelength $\lambda=1064$\,nm) along the horizontal $x$ and
$y$ directions provided a square lattice with period of $a_{\rm
lat}=\lambda/2=532$\,nm (for details see Ref.\,\cite{Sherson:2010}).
Starting from a Bose-Einstein condensate we raised the potential in
the $x$ and $y$ lattice axes within 75\,ms to values of
$V_{x,y}=23(2)\,E_{r}$ (the number in parenthesis denotes the
$1\sigma$ uncertainty of the last digit), where  $E_{r}=h^2/(2m
\lambda^2)$ is the recoil energy, and $m$ denotes the atomic mass of
$^{87}$Rb. In this way, the interaction energy of the particles with
respect to their kinetic energy was increased such that the system
undergoes a quantum phase transition to a Mott insulating
state \cite{Fisher:1989,Jaksch:1998,Greiner:2002a}. Due to the
external harmonic confinement, the Mott insulator exhibits a
shell structure with fixed integer atom numbers increasing in a
step-like manner from the outer regions of the system to the inner
core \cite{Foelling:2006,Campbell:2006}. The initial state for all
experiments presented in this paper was a single shell with only one
atom per lattice site, which in our system was realised for atom
numbers smaller than $\sim$\,400.

We detected the atoms using fluorescence imaging via a high-resolution
microscope objective with numerical aperture of
$\mbox{NA}=0.68$. An optical molasses induced fluorescence light and
simultaneously laser-cooled the atoms. Light-assisted collisions
lead to rapid losses of atom pairs, such that we only detected the
atom number modulo two on each lattice
site \cite{Bakr:2010,Sherson:2010}. With about 5,000 collected
photons per atom, we identified individual atoms in the lattice with
an excellent signal-to-noise ratio. Even in the regions of high
atomic density, we determined the presence
or absence of an atom for each lattice site with $> 99.5\%$ fidelity using a special
reconstruction algorithm \cite{Sherson:2010}.
\section*{Addressing single lattice sites}
%
%
%
\begin{figure*}
    \begin{center}
        \includegraphics[width=\textwidth]{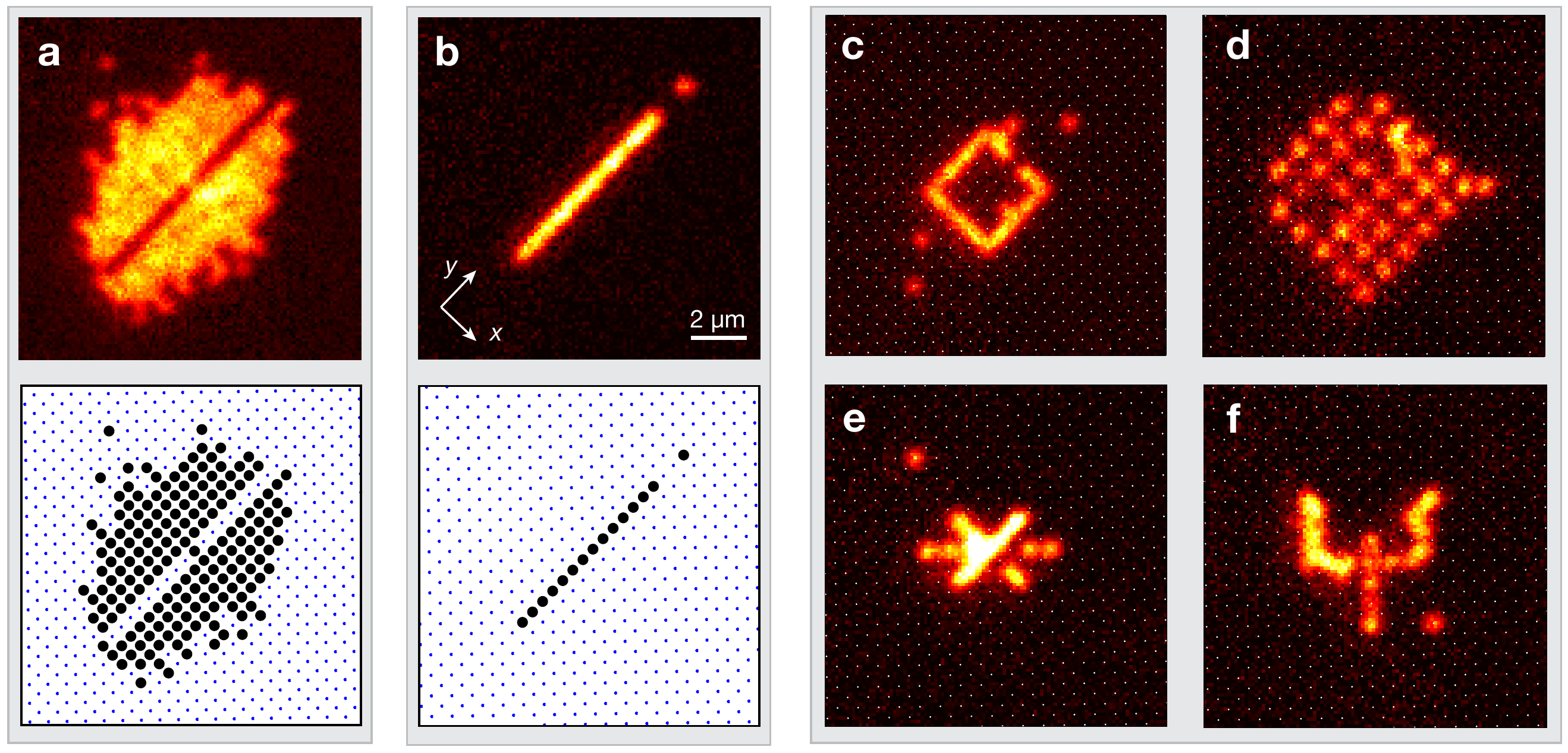}
    \end{center}
    \vspace{-0.5cm}
    \caption{{\bf Single-site addressing.}
    {\bf a,} Experimentally obtained fluorescence image of a Mott insulator with unity filling in which the spin of selected atoms was flipped from $\ket{0}$ to $\ket{1}$  using our single-site addressing scheme. Atoms in state $\ket{1}$ were removed by a resonant laser pulse before detection. The lower part shows the reconstructed atom number distribution on the lattice. Each circle indicates a single atom, the points mark the lattice sites. {\bf b-f}, Same  as {\bf a}, but a global microwave sweep exchanged the population in $\ket{0}$ and $\ket{1}$, such that only the addressed atoms were observed. The line in {\bf b} shows 14 atoms on neighbouring sites, the images {\bf c-f} contain 29, 35, 18 and 23 atoms, respectively. The single isolated atoms in {\bf b,e,f} were placed intentionally to allow for the correct determination of the lattice phase for the feedback on the addressing beam position.
 \label{fig:AddressingExamples}}
\end{figure*}
In order to address the atoms in the lattice, we used an off-resonant laser beam  focused by the high-resolution imaging system onto individual lattice sites (Fig.\,\ref{fig:schematic}). The laser beam causes a differential light shift of the two relevant hyperfine levels $\ket{0} \equiv \ket{F=1, m_F=-1}$ and $\ket{1} \equiv \ket{F=2, m_F=-2}$ and tunes the addressed atom into resonance with an external microwave field at $\sim$\,6.8 \,GHz. The $\sigma^-$-polarized addressing beam had a wavelength of $787.55$\,nm, between the $D_1$ and $D_2$ lines, in order to obtain a large differential light shift between the two hyperfine levels. For perfect $\sigma^-$-polarization, this `magic' wavelength generates a light shift only for state $\ket{1}$, and leaves state $\ket{0}$ unaffected. The beam had a diameter of $\sim 600$\,nm full-width at half-maximum (FWHM) and could be moved in the object plane over the entire field of view by changing its angle of incidence into the microscope objective with a two-axis piezo mirror. We were able to position the beam  with an accuracy better than $0.1\,\alat$ using an independent calibration measurement of its position together with a feedback that tracks the slowly varying lattice phases (see Appendix).

If the addressing laser beam is perfectly centered onto one lattice site (see Fig.\,\ref{fig:schematic}b), the differential light shift is $\Delta_{\rm LS}/(2\pi)\simeq-70$\,kHz, whereas a neighbouring atom only experiences 10\% of the peak intensity. The resulting difference in light shifts can be well resolved spectrally by our microwave pulses. In order to flip the spin, we performed Landau-Zener sweeps (see Appendix) of $\sigma_{\rm MW}/(2\pi)=60$\,kHz width and 20\,ms duration yielding a near flat-top frequency spectrum with a maximum population transfer efficiency of $\sim 95\%$.

As a first experiment, we sequentially flipped the spin of the atoms
at selected lattice sites  in our Mott insulator with unity filling
and spin state $\ket{0}$ (Fig.~\ref{fig:AddressingExamples}a). The
lattice depths were first changed to  $V_x=56\,E_{r}$,
$V_y=90\,E_{r}$ and $V_z=70\,E_{r}$ in order to completely suppress
tunnelling even when the addressing beam locally perturbs the lattice
potential. For each lattice site, we then switched on the addressing
laser beam with an s-shaped ramp within 2.5\,ms, which is adiabatic
with respect to the on-site oscillation frequency of $\sim30\,$kHz.
Subsequently, a microwave pulse with the parameters described above
produced spin-flips from $\ket{0}$ to $\ket{1}$. The addressing
laser was switched off again within 2.5\,ms, before its position was
changed in 5\,ms to address the next lattice site. For the image of
Fig.\,\ref{fig:AddressingExamples}a, this procedure was repeated 16
times in order to flip the spins along a line. Finally, a 5\,ms `push-out' laser pulse, resonant
with the $F=2$ to $F'=3$ transition, removed all addressed atoms in
state $\ket{1}$.  In order to reveal only the spin-flipped atoms, the spin states of all atoms
were flipped by a global microwave sweep before the push-out laser
was applied
(Fig.\,\ref{fig:AddressingExamples}c,d). However, due to the finite transfer efficiency of the
global sweep, some atoms remaining in state $\ket{0}$ were clearly
visible in addition to the addressed ones. To avoid this problem when detecting
the addressed atoms, we initially transferred the whole sample to
state $\ket{1}$ by a microwave sweep and then shone in a repumping
laser that completely depopulated state $\ket{0}$. Then, we used our
addressing scheme to transfer selected atoms back to $\ket{0}$ and
subsequently pushed out the atoms in $\ket{1}$, yielding typical
images as shown in Fig.\,\ref{fig:AddressingExamples}b,e,f.

%
\section*{Spin-flip fidelity}
%
%
%
\begin{figure}[!t]
    \begin{center}
        \includegraphics[width=\columnwidth]{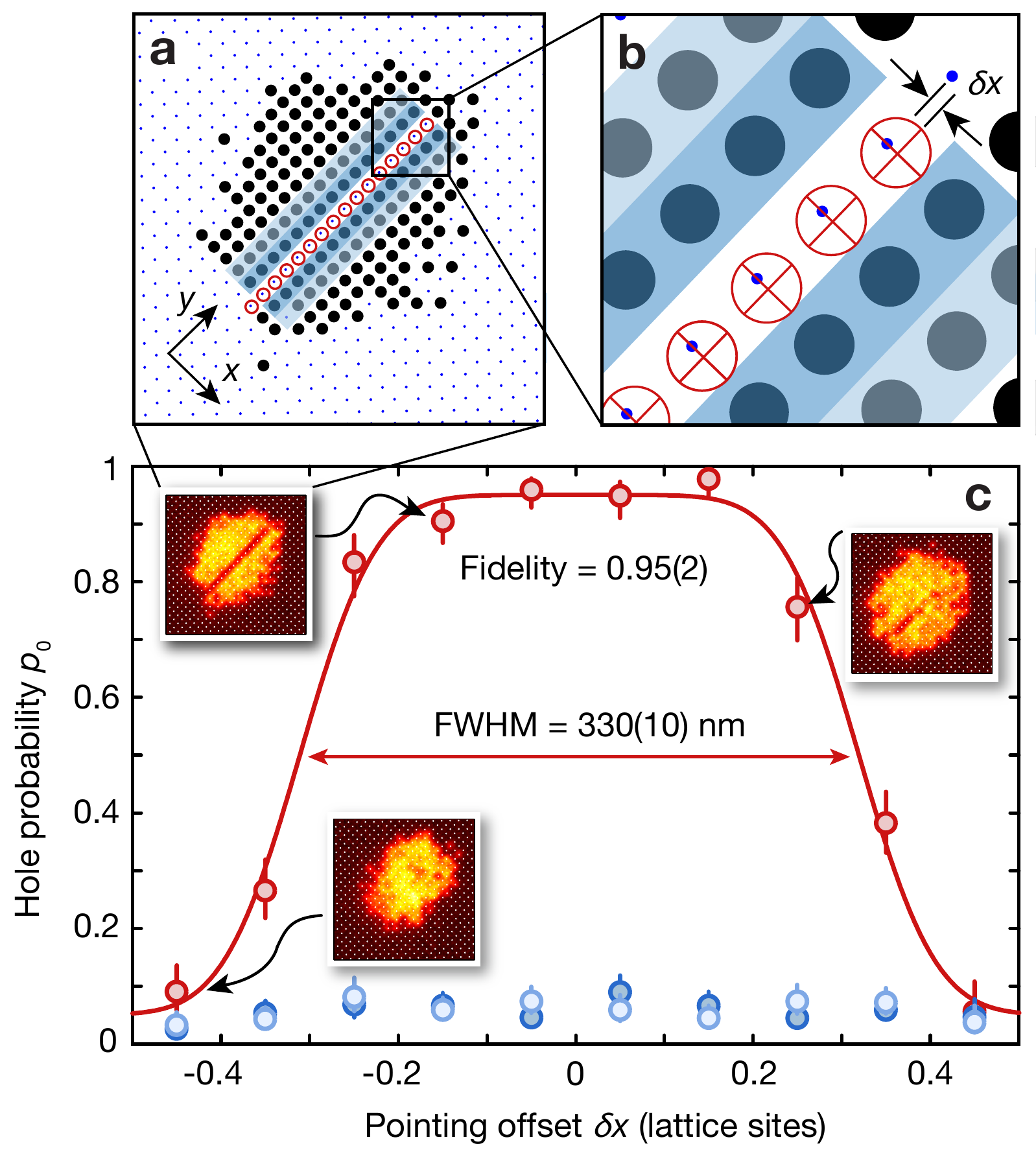}
      \vspace{-0.5cm}
    \end{center}
    \caption{{\bf Addressing fidelity.}
The spin-flip probability was measured by sequentially addressing a
series of 16 neighbouring sites along the $y$ lattice axis (red
circles in {\bf a}) in a Mott insulator with unity filling. The red
data points in {\bf c} show the resulting hole probability
$p_0(\delta x)$ as a function of the pointing offset $\delta x$, as
defined in {\bf b}. Each point was obtained by averaging over $4-7$
pictures (total $50-100$ addressed lattice sites), taking only those
sites into account which lie well within a Mott shell with unity filling.
The displayed error bars show the $1\sigma$ statistical uncertainty,
given by the Clopper-Pearson confidence limits. The data was fitted
by a flat-top model function (see Appendix) and yields a full-width
at half-maximum $\sigma_a = 330(20)$\,nm, an edge sharpness of
$\sigma_s=50(10)$\,nm, and a peak fidelity of 95(2)\%. The offset
was fixed at the 6(2)\% probability  of thermally activated holes
as deduced from the neighbouring and next neighbouring sites (blue
shaded regions in {\bf a,b} and blue points in {\bf
c}).\label{fig:Fidelity}}
\end{figure}

We quantified the success rate of our addressing scheme by again producing a series of spin-flips along the $y$ lattice axis in a Mott insulator with unity filling (see Fig.\,\ref{fig:Fidelity}). The experimental procedure was the same as described above for the realisation of Fig.\,\ref{fig:AddressingExamples}a, in which the addressed sites were detected as empty sites. From the reconstructed atom number distribution (Fig.\,\ref{fig:Fidelity}a), we determined the probability $p_0(\delta x)$ of finding an empty site as a function of the pointing offset $\delta x$ between the addressing beam and the center of the lattice site (see Fig.\,\ref{fig:Fidelity}b). We also investigated the effect of the addressing on neighbouring atoms, which ideally should remain unaffected. For this purpose, we monitored the probability of finding a hole at the sites next to the addressed ones (dark blue regions in Fig.\,\ref{fig:Fidelity}a,b and points in Fig.\,\ref{fig:Fidelity}c). In order to distinguish accidentally flipped neighbouring atoms from holes that originate from thermal excitations of the initial Mott insulator \cite{Sherson:2010}, we also monitored the probability of finding a hole at  the second next neighbours (light blue regions and points in Fig.\,\ref{fig:Fidelity}). As both yielded the same hole probability of 6(2)\%, we attribute all holes to thermal excitations and conclude that the probability of addressing a neighbouring atom is indiscernibly small. We fitted the hole probability $p_0(\delta x)$ of the addressed site with a flat-top model function (see Appendix), keeping the offset fixed at the thermal contribution of 6\%. From the fit, we derived a spin-flip fidelity of 95(2)\%, a full-width at half-maximum of $\sigma_a=330(10)$\,nm and an edge sharpness of $\sigma_s=50(10)$\,nm (Fig.\,\ref{fig:Fidelity}c). These values correspond to 60\% and 10\% of the addressing beam diameter, demonstrating that our method reaches sub-diffraction-limited resolution, well below the lattice spacing.

The observed maximum spin-flip fidelity is currently limited by the population transfer
efficiency of our microwave sweep. The edge sharpness $\sigma_s$
originates from the beam pointing error of $\lesssim 0.1\,\alat$ and
from variations in the magnetic bias field. The latter causes
frequency fluctuations of $\sim 5\,$kHz, which translate into an
effective pointing error of 0.05$\,\alat$ at the maximum slope of
the addressing beam profile. The resolution $\sigma_a$
could in principle be further reduced by a narrower microwave sweep, at the
cost of a larger sensitivity to the magnetic field fluctuations. A
larger addressing beam power would reduce this sensitivity, but we
observed that this deformed the lattice potential due to the imperfect $\sigma^{-}$-polarization, allowing neighbouring atoms to tunnel to the addressed sites.

%
\section*{Coherent tunneling dynamics}
%
%
%
\begin{figure*}
    \begin{center}
        \includegraphics[width=\textwidth]{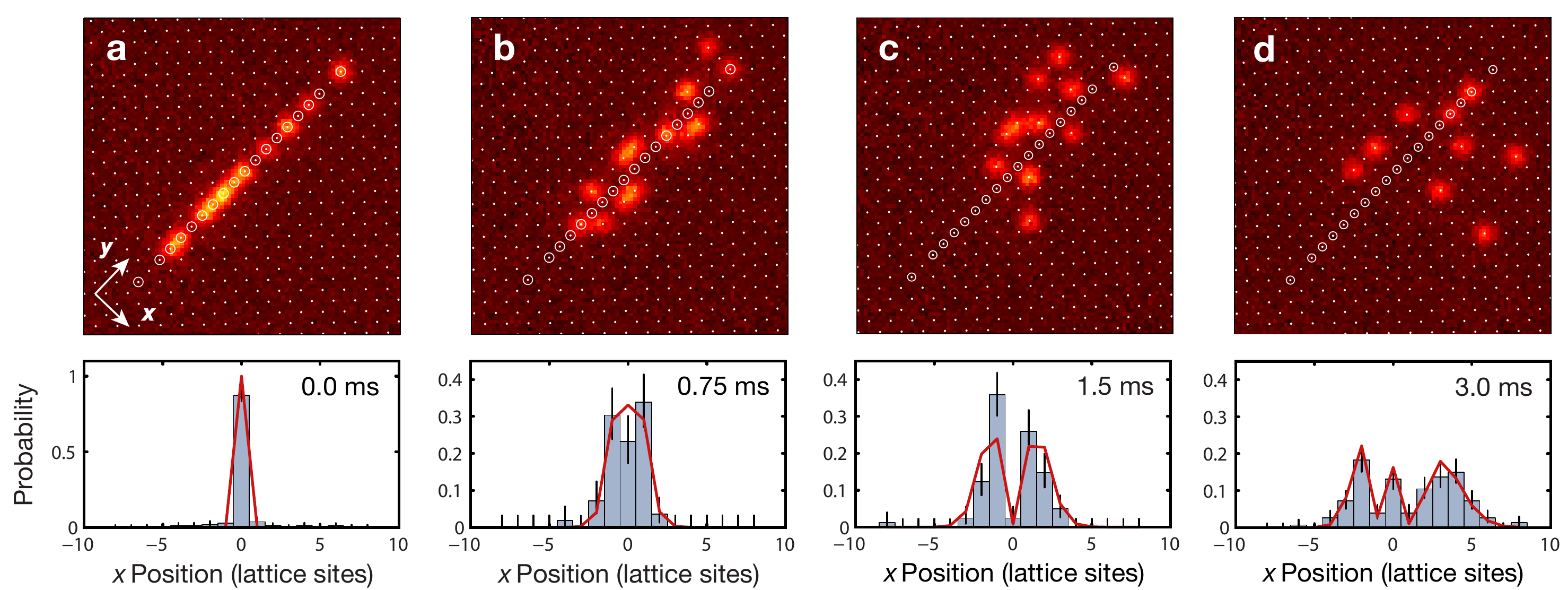}
    \end{center}
    \caption{{{\bf Ground state tunnelling dynamics.}
{\bf a,} Atoms were prepared in a single line along the $y$ direction before the lattice along the $x$ axis was lowered, allowing the atoms to tunnel in this direction ({\bf b-d}). The top row shows snapshots of the atomic distribution after different hold times. White circles indicate the lattice sites at which the atoms were prepared (not all sites initially contained an atom). The bottom row shows the respective position distribution obtained from an average over $10-20$ of such pictures, the error bars give the $1\sigma$ statistical uncertainty. A single fit to all distributions recorded at different hold times (red curve) yields a tunnelling coupling of $J^{(0)}/\hbar=940(20)$\,Hz, a trap frequency of  $\omega_{\rm trap}/(2\pi)=103(4)\,$Hz and a trap offset of $x_{\rm offs}=-6.3(6)\,\alat$.}\label{fig:HorseRace}}
\end{figure*}
The preparation of an arbitrary atom distribution opens up new
possibilities for exploring coherent quantum dynamics at the
single-atom level. As an example, we studied the
tunnelling dynamics in a one-dimensional lattice (see
Fig.\,\ref{fig:HorseRace}) which allowed us to determine how much our addressing scheme affects the vibrational state of the atoms. We started by preparing a single line of
up to 18 atoms along the $y$ direction before we lowered the lattice
along the $x$ direction to $V_x=5.0(5)E_r$ within $200\,\mu$s. At the same time,
the other lattices were lowered to $V_y=30\,E_r$ and $V_z=23\,E_r$,
which reduced the external confinement along the $x$ direction, but still suppressed
tunnelling in the $y$ and $z$ directions. After a varying hold time $t$,
allowing the atoms to tunnel along $x$, the atomic distribution was
frozen by a rapid $100\,\mu$s ramp of all lattice axes to $70\,E_r$. By averaging the resulting atomic distribution along the $y$
direction and repeating the experiment several times, we obtained
the probability distribution of finding an atom at the different
lattice sites (Fig.\,\ref{fig:HorseRace}, bottom row).

This probability distribution samples the single-atom wave function
after a coherent tunnelling evolution. We observed how the wave function expands in the
lattice and how the interference of different paths leads to
distinct maxima and minima in the distribution, leaving for example almost no
atoms at the initial position after a single tunnelling time
(Fig.\,\ref{fig:HorseRace}c). This behaviour differs markedly from the
evolution in free space, where a Gaussian wave packet disperses
without changing its shape, always preserving a maximum probability in the center.
For longer hold times, an asymmetry
in the spatial distribution becomes apparent
(Fig.\,\ref{fig:HorseRace}d), which originates from an offset
between the bottom of the external harmonic confinement and the
initial position of the atoms. We describe the observed tunnelling dynamics by a simple Hamiltonian including the tunnel coupling
$J^{(0)}$ between two neighbouring sites and an external harmonic
confinement, parameterized by the trap frequency $\omega_{\rm
trap}$, and the position offset $x_{\rm offs}$ (see Appendix). A
single fit to all probability distributions recorded at different
hold times yields $J^{(0)}/\hbar=940(20)$\,Hz, $\omega_{\rm
trap}/(2\pi)=103(4)$\,Hz and $x_{\rm offs}=-6.3(6)\,\alat$. This is
in agreement with the trap frequency $\omega_{\rm
trap}/(2\pi)=107(2)$\,Hz obtained from an independent measurement
via excitation of the dipole mode without the $x$ lattice, whose contribution to the external confinement is negligible compared to the other two axes. From $J^{(0)}$, we calculated a
lattice depth of $V_x=4.6(1)\,E_r$, which agrees with an independent calibration via
parametric heating. The expansion of the wave packet can also be
understood by writing the initial localized wave function as a
superposition of all Bloch waves of quasi-momentum $\hbar
q$, with $-\pi/\alat \leq q \leq \pi/\alat$. To each
quasi-momentum $\hbar q$, one can assign a velocity
$v_{q}= \frac{1}{\hbar}\frac{\partial E}{\partial
q}$, determined by the dispersion relation
$E(q)=-2 J^{(0)} \cos(q \alat)$ of the lowest
band. The edges of the wave packet propagate with the largest
occurring velocity $v_{\rm max}=2 J^{(0)}
\alat/\hbar=1.88(4)\,\alat/$ms, in agreement with our data.

Our measurements constitute the first observation of the ground state tunnelling dynamics of massive particles on a lattice with single-site resolution. A similar continuous-time quantum walk has been demonstrated with photons in an array of evanescently coupled photonic waveguides \cite{Perets:2008}. For massive particles, a discrete quantum walk of single atoms has been observed using a sequence of spin manipulations and spin-dependent transports in an optical lattice \cite{Karski:2009b} and also with trapped ions \cite{Zaehringer:2010}. Without single-particle and single-site resolution, a continuous-time quantum walk in the ground state has been observed for ultracold fermionic atoms by measuring their ballistic expansion in a weak lattice \cite{Schneider:2010}.

%
%
\begin{figure}[!t]
    \begin{center}
        \includegraphics[width=\columnwidth]{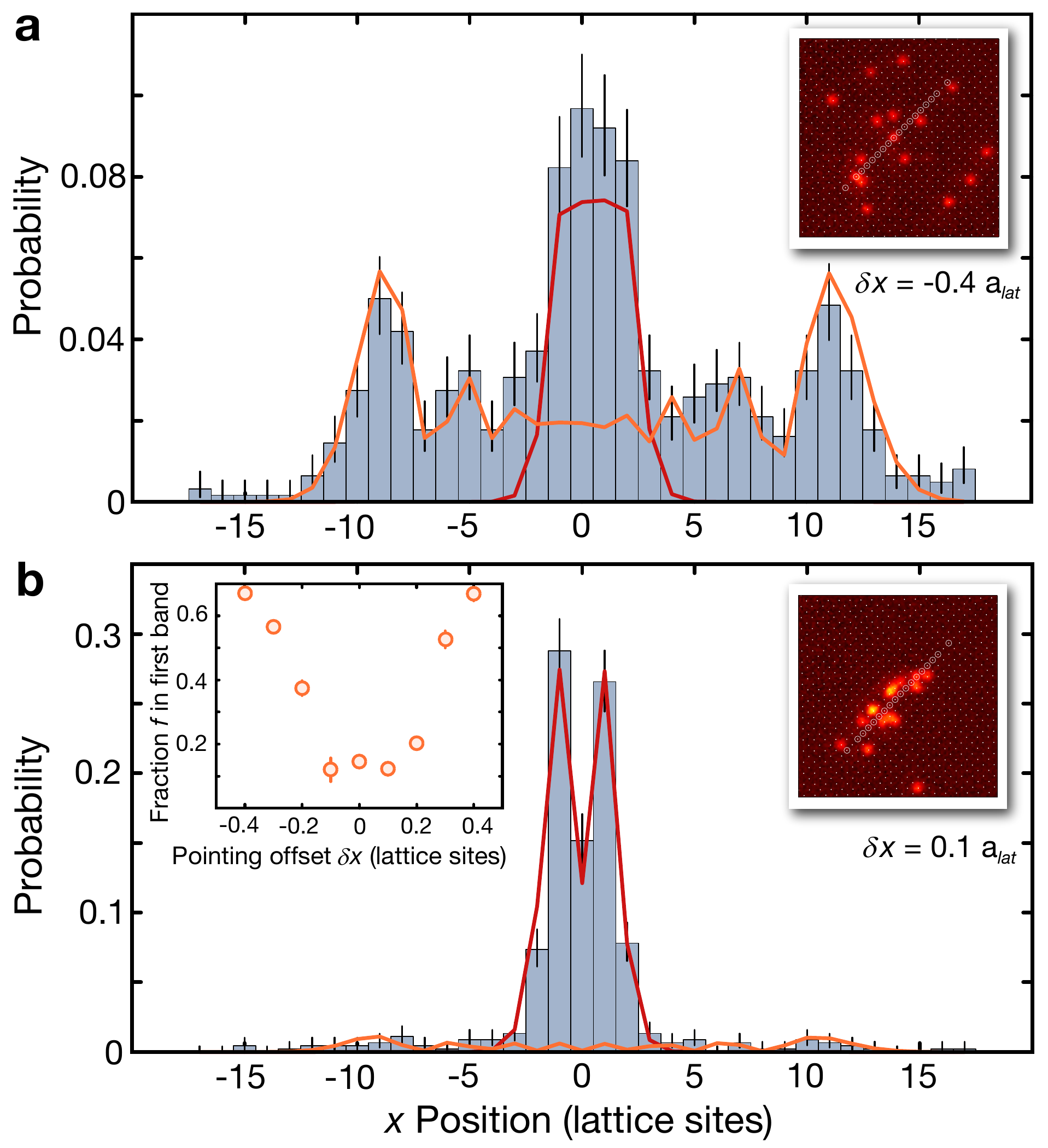}
    \end{center}
    \caption{{\bf Tunnelling dynamics of the first excited band.}
Some atoms were excited to the first band by a pointing offset $\delta x$ of the addressing beam.
{\bf a} and {\bf b} show the atomic position distribution after 1\,ms tunnelling time for $\delta x=-0.4\,\alat$ and $\delta x=0.1\,\alat$, respectively. We fitted the data with a model that includes atoms in the zeroth band (red line) and a fraction $f$ in the first band (orange line) (see Appendix). The right insets of {\bf a, b} show corresponding original images. The left inset of {\bf b} shows $f$ versus $\delta x$ with a broad minimum of $f=13(2)$\%, indicating that most of the atoms are left in the ground state. \label{fig:HotTunneling}}
\end{figure}

In a second tunnelling experiment, we observed the faster dynamics of
atoms in the first excited band (see Fig.\,\ref{fig:HotTunneling}).
For this, we deliberately excited the atoms by introducing a pointing offset $\delta x$ of
the addressing beam, which caused a shift of the potential wells
during the switch-on. We repeated the same tunnelling experiment as
above with a hold time of $t=1$\,ms for different pointing
offsets $\delta x$.  For a small pointing offset ($\delta x = 0.1\,\alat$ in Fig.\,\ref{fig:HotTunneling}b) we observed a narrow distribution, compared to a much broader one for a large offset ($\delta x = -0.4\,\alat$ in Fig.\,\ref{fig:HotTunneling}a). We attribute this to different fractions $f$ of atoms in the first band which is characterized by the higher tunnelling rate $J^{(1)}$.
We fitted the distribution of
Fig.\,\ref{fig:HotTunneling}a to a two-band model (see Appendix) and found
$J^{(1)}/\hbar=6.22(6)$\,kHz. This is in
excellent agreement with the expected value of
$J^{(1)}/\hbar=6.14(6)$\,kHz from a band structure calculation in
which we used $J^{(0)}$ as an input parameter to calculate the lattice depth. Our
measurement of the fraction of excited atoms $f$ as a function of
the pointing offset $\delta x$ (inset in
Fig.\,\ref{fig:HotTunneling}b) shows that the atoms are strongly
heated for large pointing offsets. By contrast, only a small
fraction of the atoms is excited to the first band for small pointing offsets $|\delta x| \leq 0.1\,\alat$, yielding a ground state population of $1-f=87(2)$\%.

%
\section*{Discussion}
In summary, we have demonstrated full two-dimensional single-site and single-atom spin control
in an optical lattice with sub-diffraction-limited spatial
resolution. Starting from a Mott insulator with unity filling, we
achieved a spin-flip fidelity of 95(2)\% with negligible influence
on the neighbouring lattice sites. Our scheme leaves most of the
atoms in the vibrational ground state. The control of
single spins in a strongly correlated many-body system on a lattice
opens many new possibilities for studying quantum dynamics and quantum
phases. Our technique will allow us to create out-of-equilibrium
states or local perturbations in order to observe the ensuing dynamics of
the many-body system, such as spin-charge
separation \cite{Recati:2003,Kleine:2008} or
spin impurity dynamics beyond Luttinger liquid theory \cite{Zvonarev:2007}. Our studies of the tunnelling
dynamics at the single-atom level can be extended to correlated particle tunnelling \cite{Winkler:2006,Foelling:2007,Peruzzo:2010}, also in higher dimensions, or to observe
transport across local impurities \cite{Micheli:2004} or potential barriers. The tunneling can also be used to circumvent the pairwise losses during the imaging \cite{Bakr:2010,Sherson:2010} by letting the atoms of a one-dimensional system spread along the perpendicular direction in order to obtain a sufficiently low density. Further prospects are the implementation of novel cooling schemes relying on
the local removal of regions with high entropy \cite{Sherson:2010b,Bernier:2009}.
The single-spin control in our large systems with several hundreds
of atoms also opens new perspectives for scalable quantum computing.
Combining single-qubit manipulation with local readout and a global
entanglement operation in a spin-dependent
lattice \cite{Jaksch:1999,Mandel:2003} would be the basis of a one-way quantum
computer \cite{Raussendorf:2001,Briegel:2009}. For the circuit model of a quantum
computer, two-qubit operations can be realised by Rydberg gates
between selected atom pairs in the
lattice \cite{Wilk:2010,Isenhower:2010}.


\section*{Acknowledgements}
 We thank Wolfgang Ketterle for stimulating discussions and
valuable ideas. We acknowledge the help of Rosa Gl\"{o}ckner and
Ralf Labouvie during the construction of the experiment. We
acknowledge funding by MPG, DFG, Stiftung Rheinland-Pfalz f\"{u}r
Innovation, Carl-Zeiss Stiftung, EU (NAMEQUAM, AQUTE, Marie Curie
Fellowships to J.F.S. and M.C.), and JSPS (Postdoctoral Fellowship
for Research Abroad to T.F.).

\bibliography{AddressingPreprint}


\section*{Appendix}
\subsection*{Calibration of the addressing beam position}
\vspace{-0.3cm}
To move the addressing laser beam in the object plane, we changed the angle of the beam entering from the reverse direction into the microscope objective using a two-axis piezo mirror. The device has an angular resolution of 5\,$\mu$rad, corresponding to a theoretical resolution in the object plane of $0.02\,\alat \simeq 10$\,nm. In order to position the addressing laser beam onto the atoms with high precision, we measured calibration functions  that translate the two control voltages of the piezo mirror into image coordinates. This calibration was performed by replacing the far detuned addressing laser beam by a near resonant molasses beam that follows the identical beam path. Using in addition the $x$ and $y$ molasses beams, we took a fluorescence image of a large thermal atom cloud in the vertical lattice alone and observed a
strongly enhanced signal at the position of the focused beam. We determined the position of this fluorescence maximum with an uncertainty of 0.2 pixels in our images, corresponding to $0.05\,\alat = 25$\,nm in the object plane. The long term drifts of the addressing beam position are on the order of $0.1\,\alat$ per hour, which we took into account by regular recalibration of the beam position.

\subsection*{Lattice phase feedback}
\vspace{-0.3cm}
In order to compensate slow phase drifts of the optical lattice, we applied a feedback on the position of the addressing beam. We determined the two lattice phases along $x$ and $y$ after each realisation of the experiment by fitting the position of isolated atoms. Averaging over the positions of typically $1-5$ isolated atoms per image allowed us to determine the  lattice phase to better than $0.05\,\alat$. For the determination of the phase, we used the lattice constant and the lattice angles determined from a fluorescence image with many isolated atoms \cite{Sherson:2010}. Since our phase drifts were slower than $0.04\,\alat$ between two successive realisations of the experiment (25\,s cycle time), we used the lattice phase from the last image to correct the addressing beam position. This feedback was done by adding appropriate offsets to the piezo control voltages.

\subsection*{Microwave sweeps}
\vspace{-0.3cm}
Our microwave sweeps are HS1-pulses
 \cite{Garwood:2001} with time-dependent Rabi frequency $\Omega(t)$ and detuning
$\delta(t)$ given by
\begin{subequations}
\begin{eqnarray}\label{eq:HS1power}
\Omega(t)&=&\Omega_{0}\ {\rm sech}\left[ \beta \left(\frac{2 t}{T_p}-1\right)\right] \\
\delta(t)&=&\frac{\sigma_{\rm MW}}{2}\ \tanh \left[ \beta \left(\frac{2
t}{T_p}-1 \right)\right],
\end{eqnarray}
\end{subequations}
where $\Omega_{0}/(2\pi)=3\,$kHz is the maximum Rabi-frequency, $\beta=5$ is a truncation factor, $T_p=20\,$ms is the pulse length,  and $\sigma_{\rm MW}/(2\pi)=60\,$kHz is the sweep width. The detuning $\delta(t)$ is measured relative to the center of the sweep at $\omega_{\rm MW}=\omega_0 - \Delta_{\rm MW}$ (see Fig.\,\ref{fig:schematic}b). Here, $\omega_0$ is the bare resonance between the two hyperfine states, including the shift of -570\,kHz due to the magnetic bias field along the $z$ direction and $\Delta_{\rm MW}/(2\pi)=-75\,$kHz denotes the offset of the sweep center.

\subsection*{Spin-flip fidelity}
\vspace{-0.3cm}
In order to determine the spin-flip
fidelity, we fitted the hole probability $p_0$ as a function of the
pointing offset $\delta x$ (see Fig.\,\ref{fig:Fidelity}b) to a
flat-top model function given by
\begin{equation}\label{eq:p0}
 p_0(\delta x)= \frac{A}{2}
    \left[\erf\left(\tfrac{\delta x+\sigma_a/2}{\sigma_s}\right)
        + \erf\left(-\tfrac{\delta x -\sigma_a/2}{\sigma_s}\right)
    \right] + B.
\end{equation}
Here, $\erf(x)=2/\sqrt{\pi} \int_0^x e^{-\tau^2} d\tau$ is the error function, $\sigma_a$ denotes the full-width at half-maximum of the flat-top profile and $\sigma_s$ the edge sharpness. We chose this model function since our HS1-pulses (see above) produce a flat-top population transfer profile, the edges of which are dominated by randomly fluctuating quantities (beam pointing and magnetic fields) following Gaussian statistics. The addressing fidelity is defined as  $F = A / (1-B)$ taking into account that the maximum hole probability $p_0^{\rm max} = A+B$ also includes holes from thermal defects. These yield a hole with probability $B$ at unsuccessfully addressed sites which occur with probability $1-F$, such that $p_0^{\rm max} = F + (1-F)B$.
\subsection*{Single-particle tunneling dynamics}
\vspace{-0.3cm}
We describe the coherent
tunneling dynamics on $k=2n+1$ lattice sites by the Hamiltonian
\begin{eqnarray}\label{e:singleBandTunneling}
 \hat{H}^{(0)}
& = &
 -J^{(0)} \sum_{i\,=\,-n}^{n}
   \left( \hat{a}_i^+ \hat{a}_{i+1} + \hat{a}_i^+ \hat{a}_{i-1}
   \right)\nonumber\\
   &&   + V_{\rm ext} \sum_{i\,=\,-n}^{n} {(i-x_{\rm offs})^2
   \hat{a}_i^+ \hat{a}_i},
\end{eqnarray}
where $J^{(0)}$ is the tunnel coupling in the lowest band,
$\hat{a}_i^+$ ($\hat{a}_i$) is the creation (annihilation)
operator for a particle at site $i$. The strength of the external
harmonic potential with trapping frequency $\omega_{\rm trap}$ is
given by $V_{\rm ext}=\frac{1}{2} m \omega_{\rm trap}^2 \alat^2$, and
$x_{\rm offs}$ describes a position offset with respect to the
bottom of the harmonic potential. The single particle wave function
and its coherent time evolution are given by
\begin{eqnarray}\label{e:timeEvolution}
    \Psi^{(0)}(t)
    &=&
    \sum_{i\,=\,-n}^{n}{c^{(0)}_i(t)\hat{a}_i^+}\!\ket{\tilde{0}}\nonumber\\
    &=& \exp \bigl( -i\hat{H}^{(0)}t/\hbar \bigr)\Psi^{(0)}(0),
\end{eqnarray}
with the initial condition
$\Psi^{(0)}(0)=\hat{a}_0^+\ket{\tilde{0}}$ and the vacuum state
$\ket{\tilde{0}}$. The resulting probability of finding the particle
at lattice site $i$ after time $t$ is
$P^{(0)}_i(t)=|c^{(0)}_i(t)|^2$.
For analyzing the data of Fig.\,\ref{fig:HorseRace}, we calculated
the time evolution for $k = 17$ lattice sites.
\subsection*{Tunneling in the first excited band}
\vspace{-0.3cm}
When including tunnelling in the first band, we assume an incoherent
sum $P^{\rm tot}(t)$  of the distributions $P^{(0)}(t)$ of the zeroth
and $P^{(1)}(t)$ of the first band as
\begin{equation}\label{e:fracFirstBand}
    P^{\rm tot}(t)=(1-f)P^{(0)}(t)+f P^{(1)}(t).
\end{equation}
The Hamiltonian $\hat{H}^{(1)}$ in the first band and the coherent
dynamics are identical to the ones of the zeroth band
(Eqs.\,\ref{e:singleBandTunneling} and \ref{e:timeEvolution}),
except for a different tunnel coupling $J^{(1)}$. When fitting
this model to our data, we kept $\omega_{\rm trap}$, $x_{\rm offs}$
and $J^{(0)}$ fixed at the values obtained from
the data displayed in Fig.\,\ref{fig:HorseRace}.
We extracted $J^{(1)}$ from the data of Fig.\,\ref{fig:HotTunneling}a and used this value to fit the results for other
pointing offsets. For the data in Fig.\,\ref{fig:HotTunneling}, the
parameters of our microwave sweep were such that also neighbouring
atoms were addressed. We took this into account by summing over two
distinct probability distributions with a second starting
position in the direction of the pointing offset.

\end{document}